\documentclass[10pt,conference]{IEEEtran}

\usepackage{graphicx}
\usepackage{xcolor}
\usepackage{listings}
\usepackage{tabularx}
\usepackage{booktabs}
\usepackage{array}
\usepackage{hyperref}
\usepackage{float}
\usepackage{multirow}
\usepackage{balance}
\usepackage{listings}
\newcolumntype{L}{>{\raggedright\arraybackslash}X}

\newcommand{\TODO}[1]{\textcolor{red}{#1}\GenericWarning{}{LaTeX Warning: TODO: #1}}\newcommand\todo\TODO

\lstdefinestyle{inline}{
   basicstyle=\small\ttfamily,
   keywordstyle=\color{black},
   backgroundcolor=\color{yellow},
   breaklines=true,
}

\lstdefinestyle{mystyle}{
    basicstyle=\footnotesize\ttfamily, % Default font
    numberstyle=\tiny, % Style of line numbers
    numbersep=5pt,% Margin between line numbers and text
    tabsize=2,  % Size of tabs
    extendedchars=true,
    breaklines=true,  % Lines will be wrapped
    keywordstyle=\color{red},
    frame=lines,
    columns=fullflexible,
    stringstyle=\color{white}\ttfamily, % Color of strings
    showspaces=false,
    showtabs=false,
    xleftmargin=5pt,
    framexleftmargin=5pt,
    framexrightmargin=5pt,
    framexbottommargin=5pt,
    showstringspaces=false,
    moredelim=**[is][\color{red}]{~}{~}, 
    emph={@kwsites/file-exists@ˆ1.1.1}, 
    emphstyle=[1]\color{red} 
}

\usepackage{xspace}
\newcommand{\dirtywater}{{\sc Dirty-Waters}\xspace}

\begin{document}
\title{Dirty-Waters: Detecting Software Supply Chain Smells}

\author{Raphina Liu,~\IEEEmembership{}
        Sofia Bobadilla,~\IEEEmembership{}
        Benoit Baudry,~\IEEEmembership{}
        Martin Monperrus.~\IEEEmembership{}% <-this % stops a space
        \\
        KTH Royal Institute of Technology
        }

\markboth{Journal of \LaTeX\ Class Files,~Vol.~14, No.~8, August~2015}%
{Shell \MakeLowercase{\textit{et al.}}: Bare Demo of IEEEtran.cls for IEEE Journals}

\maketitle

\begin{abstract}
Using open-source dependencies is essential in modern software development. However, this practice implies significant trust in third-party code, while there is little support for developers to assess this trust. As a consequence, attacks have been increasingly occurring through third-party dependencies. These are called software supply chain attacks. In this paper, we target the problem of projects that use dependencies while unaware of the potential risks posed by their software supply chain.
We define the novel concept of software supply chain smell and present \dirtywater, a novel tool for detecting software supply chain smells.
We evaluate \dirtywater on three JavaScript projects across nine versions and demonstrate the prevalence of all proposed software supply chain smells. Not only are there smells in all projects, but there are many of them, which immediately reveal potential risks and provide clear indicators for developers to act on the security of their supply chain.

% \todo{add a sentence about the importance of the results: not only there are smells in all projects, but there are many of them, which immediately reveals potential risks and provides clear indicators for devs to act upon the security of their supply chain}
A video demonstrating \dirtywater is available at: \href{http://l.4open.science/dirty-waters-demo}{http://l.4open.science/dirty-waters-demo}.

% https://github.com/4open-science/l
% \todo{what's this last url?}
\end{abstract}

% \todo{report}
%\begin{IEEEkeywords}
%IEEE, IEEEtran, journal, \LaTeX, paper, template.
%\end{IEEEkeywords}

% \fi
\IEEEpeerreviewmaketitle

\section{Introduction}
% for IEEE journal papers produced under \LaTeX\ using
% IEEEtran.cls version 1.8b and later.

% intro to ssc

Using third-party libraries is a common practice in software development as it reduces development costs by avoiding “reinventing the wheel”~\cite{ohmSoKPracticalDetection2023}. Despite the benefits, this practice also means putting plenty of trust in external parties. For example, installing an average NPM package implicitly means trusting 79 transitive packages and 39 maintainers~\cite{zimmermannSmallworldHighRisks2019}.
Similarly worrisome, 93\% of crypto misusages is introduced through third-party code~\cite{wickertPythonCryptoMisuses2021} and many projects unknowingly depend on vulnerable packages\cite{drososBloatPythonsScales2024}.
% intro to ssc attack
The interdependencies among packages form a complex software supply chain, whose intricate relationships make it susceptible to so called software supply chain attacks that can cause widespread impact \cite{ohmSoKPracticalDetection2023}. 

% intro to ssc defense
% \todo{add references for concepts and techniques introduced in this paragraph}
In the landscape of software supply chain analysis, there are different approaches.
First, software composition analysis (SCA) tools\cite{dietrichSecurityBlindSpots2023} have been proposed to scan dependencies and match them to an vulnerability database. 
Second, \href{https://www.ntia.doc.gov/files/ntia/publications/sbom_minimum_elements_report.pdf}{Software Bill of Materials} (SBoM) is a formal, machine-readable inventory of software components, collects information about those components, and their hierarchical relationships.
Besides, security frameworks and models such as in-toto for integrity, \href{https://slsa.dev/}{SLSA} for incremental guidelines, have been developed for securing software supply chain security. 
Yet, to our knowledge, there is no tool that highlights the most problematic dependencies in the software supply chain of a project. 
The reason is that this requires analyzing information from multiple sources, esp. the package registries and the source code repositories. 

% intro to dirty-waters
In this paper, we present \dirtywater, a novel tool designed to detect ``software supply chain smells''. Extending the concept of code smells\cite{lacerdaCodeSmellsRefactoring2020},
% code one line definition
a software supply chain smell is a
package which matches specific bad patterns that indicate potential security issues, today or to come in the future. 
\dirtywater statically analyzes dependency files,  
package registries, and GitHub repositories to detect supply chain smells. \dirtywater generates user-friendly reports for stakeholders to make sense of the quality of their software supply chains. 

% paragraph about implementation
% \todo{NPM for XX code smells, list smells} 
\dirtywater introduces five novel software ware supply chain smells: 1) Inaccessible source code link 2) Inaccessible tag 3) Using deprecated packages 4) Using forked packages 5) Missing provenance information. It implements those smells in the context of Javascript and NPM.

% how we evaluate
To evaluate the effectiveness of \dirtywater, we apply it to three real-world and complex open-source projects.
% results and implications of the  evaluate
Our results indicate that the software supply chain smells found by \dirtywater are significantly present in all three projects and present a security risk to be addressed.

To sum up, our contributions are:
\begin{itemize}
    \item The concept of software supply chain smells, essential to improve software supply chain security;
    \item \dirtywater, a publicly available tool to automatically detect software supply chain smells in NPM;
    \item Original empirical results demonstrating the relevance of the proposed smells in practice.
    
\end{itemize}

\section{Software Supply Chain Smells}

% \todo{add 2 or 3 sentences to introduce the section}

This section provides an overview of software supply chain concepts. It defines the novel concept of ``software supply chain smell'' and discusses each type of smell handled by \dirtywater.

\subsection{Background}
% quick background and definition of SCC

\textbf{Source Code Repository.} A source code repository is a place to store and maintain a project's source
code~\cite{vuUsingSourceCode2020}. It is associated with a version control system. Popular source code repositories include \href{https://github.com/}{GitHub}, \href{https://about.gitlab.com/}{GitLab}, and \href{https://bitbucket.org/product/}{Bitbucket}. 
In theory, all software libraries have an associated source code repository.

\textbf{Package Registry.} A package registry supports developers for publishing (producer) and retrieving (consumer) libraries. It is typically custom-designed for a particular programming stack, such as NPM for Javascript, Maven Central for Java, or PyPI for Python~\cite{ladisaSoKTaxonomyAttacks2023a}. It serves as a trusted, centralized entity for package binaries and bundles.

\begin{table*}
\caption{Software Supply Chain Smells in \dirtywater}
\label{tab:ssc-smells}

\begin{tabularx}{\textwidth}{@{}l p{3.8 cm} p{3.7 cm} L@{}}
\toprule
\textbf{ID} & \textbf{Smell} & \textbf{Infomation Source} & \textbf{Description} \\
\midrule
S1 & Inaccessible Source Code Link & Package Registry & Unable to access source code repository from package metadata \\

S2 & Inaccessible Release Tag & Source Repo & Unable to trace exact source code from a release package   \\

S3 & Using Deprecated Package & Package Registry\& Source Repo & Risk to be attacked from known vulnerabilities in supply chain \\

S4 & Using Forked Package & Source Repo & Potential risk of a fork attack\\

S5 & Missing Provenance Information & Package Registry & Unable to guarantee that the package construction has not been tampered with \\
\bottomrule
\end{tabularx}
\end{table*}

\textbf{Software Supply Chain.} A software supply chain is the set of libraries, tools, and applications used to develop, build, and publish a software artifact~\cite{ohmSoKPracticalDetection2023}. The definition is recursive, the software supply chain is the transitive closure of all software supply chains of sub-components of an application.
Software supply chain attacks rely on the insertion of malicious code into components of an application's software supply chain, in order to eventually let the malicious code propagate to the target application~\cite{ladisaSoKTaxonomyAttacks2023a}. 
All consumers of dependencies need to understand and minimize the attack surface through software supply chain attacks, this is what \dirtywater contributes to.

\subsection{Smell Types}

%only the single version smells
%do not mention differential analysis

% \todo{short one paragraph intro about the concept of 'smell', refering to previous work on code or design smells}

In the software engineering community, the well-accepted concept of code smell refers to characteristics in the source code that indicate a potentially deeper problem, or will trigger one in the future.

Extending this concept, we define the concept of software supply chain smells.
A \textbf{software supply chain smell} is a package that matches specific patterns that indicate potential security issues, today or to come in the future. 

In this paper, we define the software supply chain smells shown in Table~\ref{tab:ssc-smells}. The ``description" section introduces the pattern and ``reason" section highlights the potential security risks associated with each pattern, explaining why these issues deserve attention.

% \todo{briefly explain what the 'reason' part of each smell description is about}

% \todo{merge description and reason, convince, change stage to information from}

\textbf{S1. Package has no source code link \& Package's source code is inaccessible.} 
    \textit{Description:} The source code repository link is not present in the official package metadata provided by the producer, or the provided link is not available anymore.    
    \textit{Reason:} Malicious actors could add malicious code to those packages, as changes to the source code would be untraceable.

\textbf{S2. Package's tag is inaccessible.}
    \textit{Description:} A best practice is to always have a release tag in Git. It is a smell when it is missing or unavailable.
    \textit{Reason:} Without access to the release tag, it becomes impossible to trace the exact code source for a given package version and to analyze it for malicious payload. This is of utmost importance for package registries of binary code such as Maven, or of bundled/minified code such as NPM.
    
\textbf{S3. Package is marked as deprecated:} 
    \textit{Description:} A package might be deprecated due to lack of maintenance but also because of known vulnerabilities. Deprecation can be indicated via a package metadata, or a source repository metadata such as being archived on GitHub. \textit{Reason:} Being dependent on a deprecated package or archived repository can expose a consumer project to  known vulnerabilities. Malicious actors have full access to this information.
    
\textbf{S4. Package's source code repository is a fork:}  \textit{Description:} The source code repository provided by the producer is a fork from an upstream repository. \textit{Reason:} Even though packages coming from a fork is sometimes legit, overlooking this can allow a malicious actor to trick consumers of famous repositories. A notable example is the recent \texttt{typosquatting} attack on GitHub~\cite{100000InfectedRepos}, which affected over 100 thousand repositories.

\textbf{S5. Package without provenance: } 
 \textit{Description:} A provenance file, aka build attestation, is a cryptographically signed metadata about a software artifact and its build. NPM provenance~\cite{GeneratingProvenanceStatements} allows developers to verify where a package was built and who published a package. Sigstore is used to sign and log this information, allowing consumers to verify the authenticity and integrity of the package.
 % \todo{talk about sigstore/rekor} 
 \textit{Reason:} A package with provenance gives substantial security of the source code origin.
    Without provenance, malicious actors could provide false source code information or hijack the build on compromised machines. 
    
    % \todo{SO: thought about giving a more positive take on the use of provenance, feel free to change it back}

\section{Dirty-Waters}
% a description of the tool, how it works, input, output, relevant implementation details, summary report

In this section, we present the design of \dirtywater, a novel tool for checking software supply chain smells that delivers human-readable reports. 
% about supply chain transparency. 
% \todo{first time we mention transparency: add one or two sentences about what this means and how it relates to smells}.
% As transparency 
The core novelty of \dirtywater is that it detects smells through the combined analysis of
1) the local dependency file in the repository,
2) the remote package registry, and
3) the source code repository.
% \todo{we need to have those three items, graphically visible in the figure}

\begin{figure}[t]
  \centering
  \includegraphics[trim= 100 100 100 0,clip,width=0.5\textwidth]{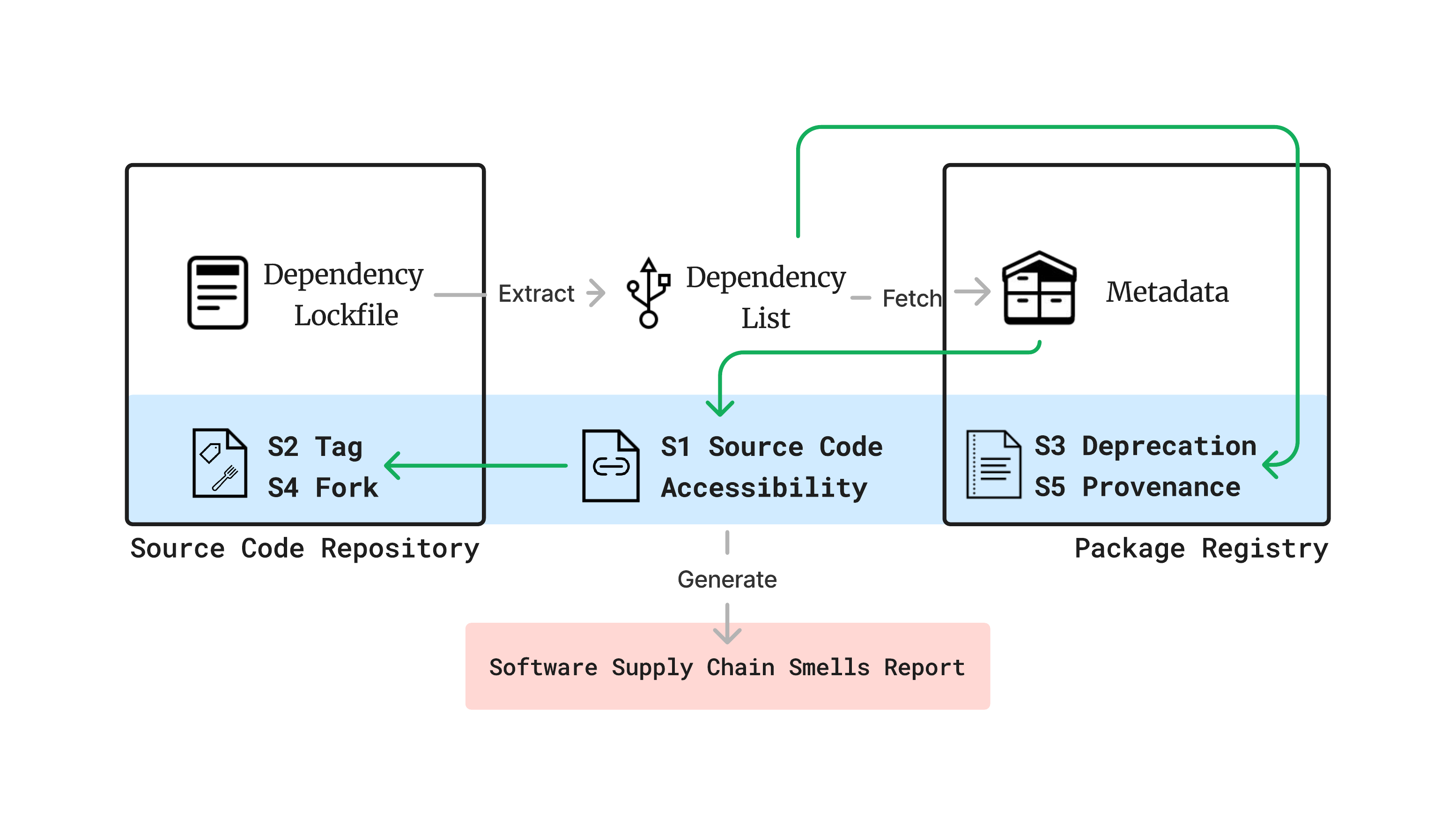}
  \vspace{-15pt}
  \caption{\dirtywater combines information from:
1) the local dependency file in the repository,
2) the remote package registry, and
3) the source code repository.}
  \label{fig:sa_flow}

\end{figure}

\subsection{Workflow}

\dirtywater receives as input:
the project name, the project version to be analyzed, and the package manager it uses.

% \todo{package metadata. source code repository and package}

The workflow, shown in Fig.~\ref{fig:sa_flow}, consists of four stages. First, the tool extracts the dependency list including both direct and transitive dependency. Next, it uses this list to find all GitHub repository links. Then, for each extracted dependency, the tool fetches metadata from 1) the package registry and 2) from the GitHub repository if any. Finally, the collected data is stored in a file that will be later used to generate the summary of software supply chain smells.

\textbf{Extract Dependency.} To extract both direct and transitive dependencies, \dirtywater analyzes the lockfile, as it provides a deterministic representation of dependencies. The specific lockfile used depends on the package manager, for example, \texttt{yarn.lock} for Yarn and \texttt{pnpm-lock.yaml} for pnpm. As shown in Listing~\ref{lst:yarn-example}, \texttt{"kwsites/file-exists@ˆ1.1.1"} is resolved to version \texttt{1.1.1} and the transitive dependency \texttt{debug "ˆ4.1.1"} is also included.

\begin{lstlisting}[firstnumber=1,float, style=mystyle, label={lst:yarn-example}, float, captionpos=b, caption=Example of yarn.lock where \dirtywater fetches the dependencies]
"@kwsites/file-exists@^1.1.1":
      version "1.1.1"
      resolved "https://registry.yarnpkg.com/@kwsites...35ef3be5b0d96faa99"
      integrity sha512-m9/...9aikZMrKPHvbpqFiw==
      dependencies:
        debug "^4.1.1"
\end{lstlisting}

% \begin{listing}
% \caption{Example of yarn.lock}
% \label{lst:example_lockfile}
% \begin{minted}[mathescape,
%                breaklines,
%                breakanywhere,
%                linenos,
%                numbersep=5pt,
%                gobble=2,
%                frame=lines,
%                fontsize=\footnotesize,
%                framesep=2mm,
%                ]{yaml}
%   "@kwsites/file-exists@^1.1.1":
%       version "1.1.1"
%       resolved "https://registry.yarnpkg.com/@kwsites/file-exists/-/file-exists-1.1.1.tgz#ad1efcac13e1987d8dbaf235ef3be5b0d96faa99"
%       integrity sha512-m9/5YGR18lIwxSFDwfE3oA7bWuq9kdau6ugN4H2rJeyhFQZcG9AgSHkQtSD15a8WvTgfz9aikZMrKPHvbpqFiw==
%       dependencies:
%         debug "^4.1.1"
% \end{minted}
% \end{listing}

\textbf{Get GitHub Repository of Dependency.} A source code link must be filled in the package metadata, as it is necessary for transparency. To get the GitHub repository link stored in package metadata, the tool executes the appropriate command based on the package manager.
For example, for the package in Listing~\ref{lst:yarn-example}, we use \lstinline[style=inline]|yarn info @kwsites/file-exists@1.1.1 repository.url| to get the repository url. This is done for detecting supply chain smell S1.

\textbf{Fetch Metadata.} \dirtywater retrieves various metadata including the deprecation status, provenance file, GitHub link accessibility, release tag, and whether the repository is a fork. For the first two, the tool queries the package registry. To check the accessibility of the GitHub link, it sends a request and verifies the HTTP response. If the GitHub link is accessible, the tool then checks the release tag’s accessibility and verifies whether the repository is a fork from data in GitHub repository. The collected data is stored in a JSON file for report generation, which we call the ``Dirty Pond''. The data from the package registry are related to supply chain smells S3 and S5 while the data related to the GitHub repository address supply chain smells S1, S2, and S4.

\begin{figure}[t]
    % \centering
    \includegraphics[width=0.65 \columnwidth]{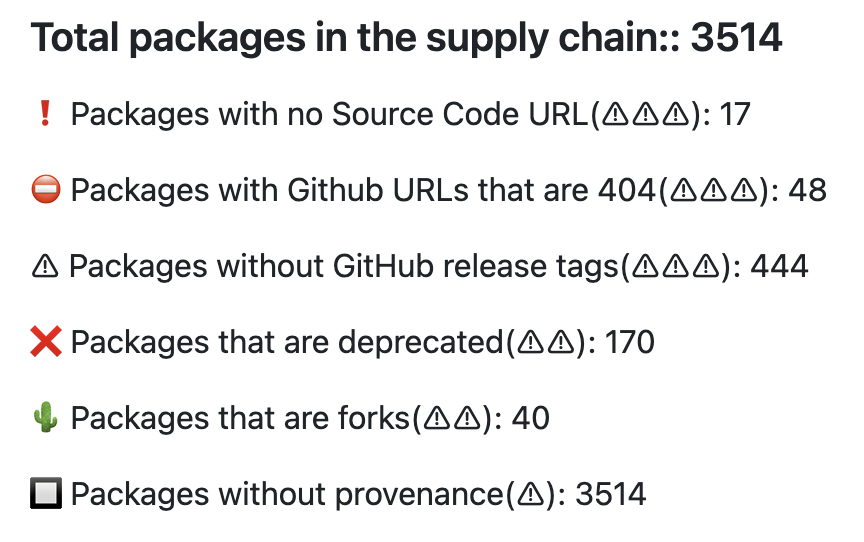}
    \caption{Software supply chain smell report for gatsby automatically generated by \dirtywater}
    \label{fig:ssc_report_example_image}
\end{figure}

\textbf{Generate Report.} Utilizing the Dirty Pond, the tool generates a comprehensive, human-readable \href{https://github.com/chains-project/dirty-waters/blob/main/example_reports/software_supply_chain_smells_report_example.md}{Markdown summary} of the analysis. As shown in Figure~\ref{fig:ssc_report_example_image},the report begins with instructions on how to interpret the results, followed by key points related to the software supply chain smells uncovered by \dirtywater.

\begin{figure}[t]
    \centering
    \includegraphics[width=0.95 \columnwidth]{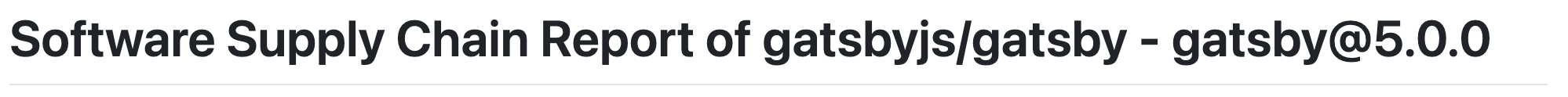}
    \vspace{-15pt}
\label{fig:title_ssc_report_example_image}
\end{figure}

\subsection{Call to Actions}

The software supply chain smells identified by \dirtywater are actionable. In the report, a ``Call to Action" section is included, to provide guidelines for developers to fix the smells:

S1 and S2: Consumers should submit a Pull Request to the dependency’s maintainer, requesting correct repository metadata and proper tagging.

S3: Confirm the maintainer's deprecation intention and double-check for alternative versions that are not deprecated.

S4: Inspect the package and its GitHub repository to verify that the fork is not malicious.

S5: Open an issue in the dependency's repository to request the inclusion of provenance and build attestation in the CI/CD pipeline. Note that, while this is still cutting-edge today, we believe that this is the future of software supply chain security.

\begin{table*}[!t]
\centering
\caption{Smells in Projects across Three Versions}
\label{tab:smells-project-experiment}
\begin{tabularx}{\textwidth}{@{}p{0.55 cm}p{3.25 cm} *{9}{c}@{}} 
\toprule
& \textbf{Project} & \multicolumn{3}{c}{\textbf{MetaMask-Extension}} & \multicolumn{3}{c}{\textbf{Webpack}} & \multicolumn{3}{c}{\textbf{Gatsby}} \\
\cmidrule(lr){3-5} \cmidrule(lr){6-8} \cmidrule(lr){9-11}
& \textbf{Version} & \textbf{v10.0.0} & \textbf{v11.0.0} & \textbf{v12.0.0} & \textbf{v5.55.0} & \textbf{v5.75.0} & \textbf{v5.95.0} & \textbf{gatsby@3.0.0} & \textbf{gatsby@4.0.0} & \textbf{gatsby@5.0.0} \\
\midrule
 & \# of Total Packages & 3597 & 3387 & 3332 & 866 & 867 & 881 & 3835 & 3615 & 3514\\
\midrule
S1 & \# with No Source Code URL & 25 & 38 & 47 & 2 & 2 & 1 & 52 & 48 & 17 \\ 
S1 &\# Source Code URL 404  & 44 & 23 & 15 & 10 & 10 & 33 & 54 & 49 & 48 \\
S2 &\# Inaccessible Tag  & 500 & 457 & 425 & 77 & 85 & 96 & 469 & 445 & 444 \\
S3 &\# of Deprecated  & 247 & 132 & 53 & 17 & 17 & 7 & 201 & 189 & 170 \\
S4 &\# from Forks  & 42 & 47 & 59 & 6 & 11 & 11 & 40 & 33 & 40 \\
S5 &\# without Provenance  & 3597 & 3384 & 3274 & 866 & 867 & 864 & 3835 & 3615 & 3514 \\

\bottomrule
\end{tabularx}
\end{table*}

\section{Experimental Evaluation}
% \subsection{intro to projects}
To evaluate the effectiveness of \dirtywater, we applied it to three large and actively maintained JavaScript projects: 
1) \href{https://github.com/MetaMask/metamask-extension}{Metamask Extension}, a popular cryptocurrency wallet extension.
2) \href{https://github.com/webpack/webpack}{Webpack}, a package used to bundle JavaScript files for usage in a browser
3) \href{https://github.com/gatsbyjs/gatsby}{Gatsby}, a react-based framework to build websites and apps.

For Metamask and Gatsby, we analyzed the latest major versions. For Webpack, we selected three versions with approximately twenty versions between each to examine software supply chain smells across a broader timeline.

% \subsection{results}
Table~\ref{tab:smells-project-experiment} shows the experimental results. The first two columns display the software supply chain smells ID and name. The next columns represent the count of packages exhibiting software supply chain smells. Each row refers to a smell.
For example, in \texttt{MetaMask-Extension@v10.0.0}, 25 packages do not have valid source code URL in the metadata.

% \todo{1) how to read the table rows/ columns/ units/ etc}
% \todo{2) concrete example row or column}

% finding 1
\emph{Large prevalence of smells. }
We see that all considered packages do have a large software supply chain, from 800+ dependencies for Webpack to more than 3500 dependencies for the other subjects.
% implications 1
It is clear that developers need tool support to understand and secure such complex software supply chains.
% finding 2
Table~\ref{tab:smells-project-experiment} reveals that all three projects suffer from the five proposed smells, demonstrating that they are not artificial or irrelevant.

% finding 3
\emph{Transparency issues. }The presence of packages with inaccessible source code links is worrisome. It means that the project's maintainers are unable to review or verify the source code of these packages they depend on, which could be part of attack vectors.
% implications
Hopefully, the Webpack developers are close to only depend on source transparent packages, which is desirable. If they can get this done, they could actually verify full transparency through a CI job, ensuring a more secure supply chain in the long term.

% finding 4
\emph{CD Immaturity. }The significant number of packages that lack corresponding tags in their GitHub repositories shows the relative immaturity of CI/CD pipelines. Dependency maintainers should always strive to create a proper tag within release scripts or continuous delivery. Dependency consumers can help out their suppliers to do that. 

% finding 5
\emph{Deprecation. }For deprecation, we observe that older versions tend to depend more on deprecated packages, reflecting the natural evolution of dependencies over time. This highlights the importance of regularly checking for software supply chain smells to keep package consumers informed of potential issues and changes, as well as using dependency bots to keep the software supply chain up-to-date with the latest versions.

% finding 6
\emph{Cutting-edge provenance. }Recall that provenance and build attestations are cutting-edge: NPM provenance was introduced in April 2023, so it is not surprising that
our results show that only a small number of packages have adopted the use of provenance. We advocate that the most security sensitive libraries and applications add provenance to their release pipelines.

% Figure
% \todo{1 wallet + webpack/webpack + react)} 3 versions each?
% \todo{https://github.com/npm/registry/blob/main/docs/download-counts.md}
% \todo{https://npmrank.net/}
% https://github.com/EvanLi/Github-Ranking/blob/master/Top100/JavaScript.md

\section{Related work}
% Code smells
Code smells increase technical debt, affecting maintenance and evolution~\cite{lacerdaCodeSmellsRefactoring2020}. Cherry et al.~\cite{cherrySMEAGOLStaticCode2024} developed a static analysis tool to detect code smells and Mello et al.~\cite{demelloRecommendationsDevelopersIdentifying2023} provide recommendations to developers to identify code smells. In this paper, we extend the concept and propose this concept of software supply chain smells.
% \todo{So: How does this relate to a software supply issue and/ or why is different, I'm looking for a transition sentence}

% https://ieeexplore.ieee.org/stamp/stamp.jsp?tp=\&arnumber=6671299

% https://martinfowler.com/bliki/CodeSmell.html

% https://www.cs.wm.edu/~denys/pubs/ICSE'15-BadSmells-CRC.pdf

% sboms
% Several standards and tools related to SBoMs have been proposed. Among them SPDX\cite{SPDXLinuxFoundation}, CycloneDX \cite{OWASPCycloneDXSoftware}, SWID\cite{NVDSWID} are the three existing formats that meet the requirements of the United States National Telecommunications and Information Administration(NTIA).

% SSC research
Previous research has discussed various aspects related to code repositories and registries, such as checking repository accessibility~\cite{tsakpinisAnalyzingAccessibilityGitHub2024}, detecting injection attacks in  artifacts\cite{vuUsingSourceCode2020}, identifying known vulnerabilities in source code~\cite{hastingsContinuousVerificationOpen2022}, analyzing project maintainer activity~\cite{zahanWhatAreWeak2022}, repository popularity~\cite{borgesUnderstandingFactorsThat2016} and examining malware in repository forks\cite{caoWhatForkFinding2022}. However, none of these works provide an actionable tool with a joint analysis of both package registry and source code repository information, as \dirtywater does.

% \todo{put cite in the right places}

% dependency smells
There is little research related to dependency smells ~\cite{jafariDependencySmellsJavaScript2022,caoBetterDependencyManagement2023}, it has explored issues like bloated dependencies, missing dependencies, and erroneous version constraints. This previous research is different in nature from \dirtywater, which analyzes the intersection between package registry and source code repository information. \dirtywater is a ready-to-use tool for providing package consumers with user-friendly reports.

% \todo{paper title from PC members}

% [ok]Dependency smells in javascript projects -> already cited

% - https://scholar.google.com/citations?hl=en&user=ryow8hIAAAAJ&view_op=list_works&sortby=pubdate
% - dependency, blockchain, defi, open source
% CrossVul: a cross-language vulnerability dataset with commit data

% [ok]Bloat beneath Python’s Scales: A Fine-Grained Inter-Project Dependency Analysis

% - https://scholar.google.com/citations?hl=en&user=WLgmKW4AAAAJ&view_op=list_works&sortby=pubdate
% - Code smell, developers, SQL
% [ok]SMEAGOL: A Static Code Smell Detector for MongoDB

% - https://scholar.google.fr/citations?hl=fr&user=8Yk4sZ4AAAAJ&view_op=list_works&sortby=pubdate
% - dependency, go, static analysis
% [ok]Python Crypto Misuses in the Wild

% - https://scholar.google.se/citations?hl=sv&user=2fwfYtQAAAAJ&view_op=list_works&sortby=pubdate
% - work with github repository
% [ok]Understanding the factors that impact the popularity of GitHub repositories

% Dependency analysis bots(SCA)
% code smells

% \todo{add: not all dependencies are equal:  https://dl.acm.org/doi/pdf/10.1145/3551349.3556896}

% https://ieeexplore.ieee.org/stamp/stamp.jsp?tp=\&arnumber=8812041

% The Seven Sins: Security Smells in Infrastructure as
% Code Scripts

% frameworks
% (\cite{intoto}, \cite{SLSA}, \cite{BSIMM}, \cite{SAMM})

% \todo{choose the 10-15 most revelant cited papers in the thesis}
% \todo{check PC members}
% \todo{replace code smell cite with PC member work}

\section{Conclusion}
Taking a closer look at the software supply chain of a software product is essential for maintaining its long-term security. In this paper, we define software supply chain smells, bad practices that emerge at the intersection of installed dependencies, package registries metadata and source code repositories.
These smells are indicators of potential security risks induced by the software supply chain. We also present \dirtywater, a novel tool that detects these smells and generates human-readable reports to enhance supply chain transparency.
We evaluate \dirtywater's effectiveness by applying it to three complex JavaScript projects — MetaMask Extension, Webpack, and Gatsby — with thousands of dependencies.
\dirtywater successfully identifies all five software supply chain smells in those projects. Future work should expand \dirtywater to support additional package ecosystems such as Java/Maven and Rust/Cargo. \dirtywater is publicly available at: \href{https://github.com/chains-project/dirty-waters}{https://github.com/chains-project/dirty-waters}.

% or




%\bibliographystyle{IEEEtran}

\balance

\bibliographystyle{IEEEtran}
\bibliography{references}

\begin{thebibliography}{10}
\providecommand{\url}[1]{#1}
\csname url@samestyle\endcsname
\providecommand{\newblock}{\relax}
\providecommand{\bibinfo}[2]{#2}
\providecommand{\BIBentrySTDinterwordspacing}{\spaceskip=0pt\relax}
\providecommand{\BIBentryALTinterwordstretchfactor}{4}
\providecommand{\BIBentryALTinterwordspacing}{\spaceskip=\fontdimen2\font plus
\BIBentryALTinterwordstretchfactor\fontdimen3\font minus \fontdimen4\font\relax}
\providecommand{\BIBforeignlanguage}[2]{{%
\expandafter\ifx\csname l@#1\endcsname\relax
\typeout{** WARNING: IEEEtran.bst: No hyphenation pattern has been}%
\typeout{** loaded for the language `#1'. Using the pattern for}%
\typeout{** the default language instead.}%
\else
\language=\csname l@#1\endcsname
\fi
#2}}
\providecommand{\BIBdecl}{\relax}
\BIBdecl

\bibitem{ohmSoKPracticalDetection2023}
\BIBentryALTinterwordspacing
M.~Ohm and C.~Stuke, ``{SoK}: {Practical} {Detection} of {Software} {Supply} {Chain} {Attacks},'' in \emph{Proceedings of the 18th {International} {Conference} on {Availability}, {Reliability} and {Security}}, ser. {ARES} '23.\hskip 1em plus 0.5em minus 0.4em\relax New York, NY, USA: Association for Computing Machinery, Aug. 2023, pp. 1--11. [Online]. Available: \url{https://dl.acm.org/doi/10.1145/3600160.3600162}
\BIBentrySTDinterwordspacing

\bibitem{zimmermannSmallworldHighRisks2019}
M.~Zimmermann, C.-A. Staicu, C.~Tenny, and M.~Pradel, ``Smallworld with high risks: a study of security threats in the npm ecosystem,'' in \emph{Proceedings of the 28th {USENIX} {Conference} on {Security} {Symposium}}, ser. {SEC}'19.\hskip 1em plus 0.5em minus 0.4em\relax USA: USENIX Association, Aug. 2019, pp. 995--1010.

\bibitem{wickertPythonCryptoMisuses2021}
\BIBentryALTinterwordspacing
A.-K. Wickert, L.~Baumgärtner, F.~Breitfelder, and M.~Mezini, ``\BIBforeignlanguage{en}{Python {Crypto} {Misuses} in the {Wild}},'' in \emph{\BIBforeignlanguage{en}{Proceedings of the 15th {ACM} / {IEEE} {International} {Symposium} on {Empirical} {Software} {Engineering} and {Measurement} ({ESEM})}}.\hskip 1em plus 0.5em minus 0.4em\relax Bari Italy: ACM, Oct. 2021, pp. 1--6. [Online]. Available: \url{https://dl.acm.org/doi/10.1145/3475716.3484195}
\BIBentrySTDinterwordspacing

\bibitem{drososBloatPythonsScales2024}
\BIBentryALTinterwordspacing
G.-P. Drosos, T.~Sotiropoulos, D.~Spinellis, and D.~Mitropoulos, ``\BIBforeignlanguage{en}{Bloat beneath {Python}’s {Scales}: {A} {Fine}-{Grained} {Inter}-{Project} {Dependency} {Analysis}},'' \emph{\BIBforeignlanguage{en}{Proceedings of the ACM on Software Engineering}}, vol.~1, no. FSE, pp. 2584--2607, Jul. 2024. [Online]. Available: \url{https://dl.acm.org/doi/10.1145/3660821}
\BIBentrySTDinterwordspacing

\bibitem{dietrichSecurityBlindSpots2023}
\BIBentryALTinterwordspacing
J.~Dietrich, S.~Rasheed, A.~Jordan, and T.~White, ``On the {Security} {Blind} {Spots} of {Software} {Composition} {Analysis},'' Oct. 2023, arXiv:2306.05534 [cs]. [Online]. Available: \url{http://arxiv.org/abs/2306.05534}
\BIBentrySTDinterwordspacing

\bibitem{lacerdaCodeSmellsRefactoring2020}
\BIBentryALTinterwordspacing
G.~Lacerda, F.~Petrillo, M.~Pimenta, and Y.~G. Gueheneuc, ``Code {Smells} and {Refactoring}: {A} {Tertiary} {Systematic} {Review} of {Challenges} and {Observations},'' \emph{Journal of Systems and Software}, vol. 167, p. 110610, Sep. 2020, arXiv:2004.10777 [cs]. [Online]. Available: \url{http://arxiv.org/abs/2004.10777}
\BIBentrySTDinterwordspacing

\bibitem{vuUsingSourceCode2020}
\BIBentryALTinterwordspacing
D.~L. Vu, I.~Pashchenko, F.~Massacci, H.~Plate, and A.~Sabetta, ``Towards {Using} {Source} {Code} {Repositories} to {Identify} {Software} {Supply} {Chain} {Attacks},'' in \emph{Proceedings of the 2020 {ACM} {SIGSAC} {Conference} on {Computer} and {Communications} {Security}}, ser. {CCS} '20.\hskip 1em plus 0.5em minus 0.4em\relax New York, NY, USA: Association for Computing Machinery, Nov. 2020, pp. 2093--2095. [Online]. Available: \url{https://doi.org/10.1145/3372297.3420015}
\BIBentrySTDinterwordspacing

\bibitem{ladisaSoKTaxonomyAttacks2023a}
\BIBentryALTinterwordspacing
P.~Ladisa, H.~Plate, M.~Martinez, and O.~Barais, ``{SoK}: {Taxonomy} of {Attacks} on {Open}-{Source} {Software} {Supply} {Chains},'' in \emph{2023 {IEEE} {Symposium} on {Security} and {Privacy} ({SP})}, May 2023, pp. 1509--1526, iSSN: 2375-1207. [Online]. Available: \url{https://ieeexplore.ieee.org/abstract/document/10179304}
\BIBentrySTDinterwordspacing

\bibitem{100000InfectedRepos}
\BIBentryALTinterwordspacing
``Over 100,000 {Infected} {Repos} {Found} on {GitHub}.'' [Online]. Available: \url{https://apiiro.com/blog/malicious-code-campaign-github-repo-confusion-attack/}
\BIBentrySTDinterwordspacing

\bibitem{GeneratingProvenanceStatements}
\BIBentryALTinterwordspacing
``\BIBforeignlanguage{en}{Generating provenance statements {\textbar} npm {Docs}}.'' [Online]. Available: \url{https://docs.npmjs.com/generating-provenance-statements}
\BIBentrySTDinterwordspacing

\bibitem{cherrySMEAGOLStaticCode2024}
\BIBentryALTinterwordspacing
B.~Cherry, C.~Nagy, M.~Lanza, and A.~Cleve, ``\BIBforeignlanguage{English}{{SMEAGOL}: {A} {Static} {Code} {Smell} {Detector} for {MongoDB}}.''\hskip 1em plus 0.5em minus 0.4em\relax IEEE Computer Society, Mar. 2024, pp. 816--820. [Online]. Available: \url{https://www.computer.org/csdl/proceedings-article/saner/2024/306600a816/1YCRoPI4vQI}
\BIBentrySTDinterwordspacing

\bibitem{demelloRecommendationsDevelopersIdentifying2023}
\BIBentryALTinterwordspacing
R.~de~Mello, R.~Oliveira, A.~Uchôa, W.~Oizumi, A.~Garcia, B.~Fonseca, and F.~de~Mello, ``Recommendations for {Developers} {Identifying} {Code} {Smells},'' \emph{IEEE Software}, vol.~40, no.~2, pp. 90--98, Mar. 2023, conference Name: IEEE Software. [Online]. Available: \url{https://ieeexplore.ieee.org/document/9904005/?arnumber=9904005}
\BIBentrySTDinterwordspacing

\bibitem{tsakpinisAnalyzingAccessibilityGitHub2024}
\BIBentryALTinterwordspacing
A.~Tsakpinis and A.~Pretschner, ``Analyzing the {Accessibility} of {GitHub} {Repositories} for {PyPI} and {NPM} {Libraries},'' Apr. 2024, arXiv:2404.17403 [cs]. [Online]. Available: \url{http://arxiv.org/abs/2404.17403}
\BIBentrySTDinterwordspacing

\bibitem{hastingsContinuousVerificationOpen2022}
\BIBentryALTinterwordspacing
T.~Hastings and K.~R. Walcott, ``Continuous {Verification} of {Open} {Source} {Components} in a {World} of {Weak} {Links},'' in \emph{2022 {IEEE} {International} {Symposium} on {Software} {Reliability} {Engineering} {Workshops} ({ISSREW})}, Oct. 2022, pp. 201--207. [Online]. Available: \url{https://ieeexplore.ieee.org/abstract/document/9985184}
\BIBentrySTDinterwordspacing

\bibitem{zahanWhatAreWeak2022}
\BIBentryALTinterwordspacing
N.~Zahan, T.~Zimmermann, P.~Godefroid, B.~Murphy, C.~Maddila, and L.~Williams, ``What are weak links in the npm supply chain?'' in \emph{Proceedings of the 44th {International} {Conference} on {Software} {Engineering}: {Software} {Engineering} in {Practice}}, ser. {ICSE}-{SEIP} '22.\hskip 1em plus 0.5em minus 0.4em\relax New York, NY, USA: Association for Computing Machinery, Oct. 2022, pp. 331--340. [Online]. Available: \url{https://dl.acm.org/doi/10.1145/3510457.3513044}
\BIBentrySTDinterwordspacing

\bibitem{borgesUnderstandingFactorsThat2016}
\BIBentryALTinterwordspacing
H.~Borges, A.~Hora, and M.~T. Valente, ``Understanding the {Factors} {That} {Impact} the {Popularity} of {GitHub} {Repositories},'' in \emph{2016 {IEEE} {International} {Conference} on {Software} {Maintenance} and {Evolution} ({ICSME})}, Oct. 2016, pp. 334--344. [Online]. Available: \url{https://ieeexplore.ieee.org/document/7816479}
\BIBentrySTDinterwordspacing

\bibitem{caoWhatForkFinding2022}
\BIBentryALTinterwordspacing
A.~Cao and B.~Dolan-Gavitt, ``\BIBforeignlanguage{en}{What the {Fork}? {Finding} and {Analyzing} {Malware} in {GitHub} {Forks}},'' in \emph{\BIBforeignlanguage{en}{Proceedings 2022 {Workshop} on {Measurements}, {Attacks}, and {Defenses} for the {Web}}}.\hskip 1em plus 0.5em minus 0.4em\relax San Diego, CA, USA: Internet Society, 2022. [Online]. Available: \url{https://www.ndss-symposium.org/wp-content/uploads/madweb2022_23001_paper.pdf}
\BIBentrySTDinterwordspacing

\bibitem{jafariDependencySmellsJavaScript2022}
\BIBentryALTinterwordspacing
A.~J. Jafari, D.~E. Costa, R.~Abdalkareem, E.~Shihab, and N.~Tsantalis, ``\BIBforeignlanguage{en}{Dependency {Smells} in {JavaScript} {Projects}},'' \emph{\BIBforeignlanguage{en}{IEEE Transactions on Software Engineering}}, vol.~48, no.~10, pp. 3790--3807, Oct. 2022. [Online]. Available: \url{https://ieeexplore.ieee.org/document/9519532/}
\BIBentrySTDinterwordspacing

\bibitem{caoBetterDependencyManagement2023}
\BIBentryALTinterwordspacing
Y.~Cao, L.~Chen, W.~Ma, Y.~Li, Y.~Zhou, and L.~Wang, ``Towards {Better} {Dependency} {Management}: {A} {First} {Look} at {Dependency} {Smells} in {Python} {Projects},'' \emph{IEEE Transactions on Software Engineering}, vol.~49, no.~4, pp. 1741--1765, Apr. 2023, conference Name: IEEE Transactions on Software Engineering. [Online]. Available: \url{https://ieeexplore.ieee.org/document/9832512/?arnumber=9832512}
\BIBentrySTDinterwordspacing

\end{thebibliography}
\addcontentsline{toc}{part}{Bibliography}


% \end{thebibliography}


% \end{IEEEbiography}


%\newpage



%\vfill


\end{document}